\documentstyle[sprocl,amstex,epsfig]{article}

\def\be{\begin{equation}}
\def\ee{\end{equation}}
\def\bea{\begin{eqnarray}}
\def\eea{\end{eqnarray}}

\begin{document}

\title{$k_\bot$-FACTORIZATION PREDICTIONS FOR $F_2^c$ AT HERA\footnote{Talk 
presented at the DIS98 workshop, Brussels, April 1998.}
}

\author{S. MUNIER}

\address{Service de Physique Th\'eorique, CEA, CE-Saclay\\
F-91191 Gif-sur-Yvette Cedex, France}


\maketitle\abstracts{High energy factorization predictions for 
$F_2^c$ are derived using BFKL resummations of leading logs for the proton 
structure functions at HERA. A theoretical non-perturbative uncertainty 
on the factorization scheme is taken into account by 
considering two different approaches
for modelling the proton. The parameters are fixed by a fit of $F_2$ 
at small $x$. The resulting predictions for $F_2^c$ 
are in agreement with the data within the present error bars.}

\section{High energy factorization}

$k_\bot$- (or high energy) factorization \cite{catani,ellis} is a QCD
factorization scheme suited for high-energy hard processes - and in particular
for deep-inelastic $e^\mp\!\!-\!\!p$ scattering 
in the small $x$-regime. This scheme takes into account the
resummation of the $\left(\alpha_s \log \frac{1}{x}\right)^n$
terms in the QCD perturbative expansion of the structure functions.

Let us formulate $k_\bot$-factorization for the leptoproduction
of a quark- antiquark pair of mass $M$ off a small size 
(``perturbative'') target characterized by its gluon distribution.
In this scheme \cite{catani}, the inclusive transverse (resp. longitudinal) 
structure functions $F_T$ ($F_L$) can be expressed as follows:
\vspace{-0.1cm}
\begin{multline}
F_{T,L}(Y,Q^2,M^2;Q_f^2) =
\frac{1}{4\pi^2\alpha_{em}}\frac{Q^2}{4{M}^2}
\int_{Q_f} d^2k_\bot\int_0^\infty dy\\
   \hat{\sigma}_{\gamma^* g,T,L}
(Y\!\!-\!\!y,q_\bot/M,k_\bot/M)\;{\mathcal F}(y,k_\bot)\ ,
\label{eqn:0}
\end{multline}
where $Q^2=-q^2$ is the virtuality of the photon, $Q_f$ the factorization 
scale and $M$ the mass of the produced quarks.
$Y$ represents the rapidity range available for the reaction. 
${\mathcal F}(y,k_\bot)$ is the unintegrated gluon distribution
\cite{catani}, which describes the probability of finding a gluon
with longitudinal momentum fraction $z=e^{-y}$ and 
two-dimensional transverse momentum $k_\bot$ in the target. 
$q_\bot$ is the photon transverse momentum.
$\hat{\sigma}_{\gamma^* g,T,L}$ is the hard cross section for 
(virtual)photon-(virtual)gluon fusion computed at order $\alpha_s\alpha_{em}$.

The final expression for the high-energy factorized structure function is
most easily expressed as an inverse Mellin transform and reads:
\begin{equation}
F_{T,L}(Y,Q^2,M^2;Q_f^2)=
e^2\!\!\int\frac{d\gamma}{2i\pi}\left(\frac{Q^2}{Q_f^2}
\right)^\gamma h_{T,L}(\gamma;M^2)
\ \frac{{\mathcal F}(Y,\gamma;Q_f^2)}{\gamma}
\label{eqn:finale}
\end{equation}
where the coefficient functions $h_{T,L}(\gamma;M^2)$ represent the
Mellin-transform of the off-shell (virtual)photon-(virtual)gluon cross section,
in an approximation in which one neglects subleading terms in energy. One can 
find their expression in \cite{catanew} and \cite{mp}.

${\mathcal F}(Y,\gamma;Q_f^2)$ is the Mellin-transform of
the unintegrated gluon distribution ${\mathcal F}(Y,k_\bot)$ with 
respect to the transverse momentum of the gluon.
It is assumed to satisfy the BFKL dynamics \cite{bfkl}.
Assuming a full factorization of the rapidity dependence, which is consistent
with an asymptotic approximation for the coefficient functions, we obtain the 
following parametrization:
\begin{equation}
\frac{{\mathcal F}(Y,\gamma;Q_f^2)}{\gamma}=
\omega(\gamma;Q_f^2)\ e^{\frac{\alpha_sN_c}{\pi}\chi(\gamma)Y}\ ,
\label{eqn:sing}
\end{equation}
where
$\chi(\gamma)=2\Psi(1)-\Psi(\gamma)-\Psi(1\!\!-\!\!\gamma)$.
The function $\omega(\gamma;Q_f^2)$ will explicitly depend
on the nature of the target, and has to be supplied by a model for
an extended target like a proton, see next section.

\section{Phenomenology}

Let us introduce the model for the proton.
Following the suggestion of ref.~\cite{nprw}, one assumes the scaling form
$\omega(\gamma;Q_f^2)=
   \omega(\gamma)(Q_f/{Q_0})^{2\gamma}$,
where $Q_0$ is a non-perturbative scale, independent of the mass $M$.
With this assumption, the overall formula (\ref{eqn:finale}) does no more depend on the
factorization scale $Q_f$.

We have shown in ref.~\cite{mp} that the behaviour of $\omega(\gamma)$ 
when $\gamma$ becomes large cannot be steeper than a polynomial. 
This constraint comes from the region where $k_\bot$ is large, 
i.e. where we expect rather a DGLAP evolution.
Taking into account this constraint, we will now focus on two definite models 
relying on different formulations of the residue function $\omega(\gamma)$ 
at small $\gamma$. 
On the one hand, the model 1, with $\omega(\gamma)=C (\mbox{constant})$
corresponds to the factorization at the gluon level: all the perturbative
content of $k_\bot$-factorization is kept. 
On the other hand, in model 2, we consider an input compensating the $1/\gamma$
pole of $h_T(\gamma;M^2)$:
$\omega(\gamma)\sim\gamma C$ at small $\gamma$.
This model corresponds to a factorization at the quark level~\cite{nikolaev}, 
and was discussed in ref.~\cite{mp}.
Both models lead to an expression for the proton structure functions depending
on three free parameters, $C$, $\alpha_s$, and $Q_0$.

We determine these parameters for both models by a fit of $F_2=F_T+F_L$ in 
their kinematical region of validity ($x\leq 10^{-2}$, moderate $Q^2$). 
Using the corresponding 103 
experimental points given by the H1 collaboration \cite{f2}, we fit
our results (\ref{eqn:finale}) with the contribution of the three light 
quarks $u,d,s$ (assumed massless) and of the charm quark (mass $M_c$). 
The $F_2$-fit for the medium mass $M_c=1.5\;\mbox{GeV}$ 
is displayed in figure 1, together with the predictions for its charm
component $F_2^c$.
For model 1, the $\chi^2$ per point is always less than $0.9$, while for model 2
it is even lower.
For model 1, the value of $Q_0$ is around 330 MeV which is a typical non perturbative
scale for the proton. The value of the effective coupling constant in
the BFKL mechanism $\alpha_s$  ($0.07$) is rather low. 
For model 2, the data for $F_2$ are also fairly well reproduced (see figure 1). 
Note that the value of $Q_0\simeq 1.2\ \mbox{ GeV}$ is substantially higher
and the effective coupling constant $\alpha_s$ is a bit larger 
($\simeq \alpha_s(M_Z)$).

The parameter free predictions \cite{mp} for $F_2^c$ obtained as an outcome
of both of the considered models are in good agreement with
ZEUS and H1 data \cite{f2c}, within the present experimental error bars, 
although model 2 predicts a significantly higher charm component.
The predictions are also comparable to the next-leading order prediction
\cite{harris} based on the GRV parton distribution set \cite{GRV}
which proves that $F_2^c$ cannot allow one to distinguish between
the two approaches.

\begin{figure}[t]
\begin{center}
\mbox{\epsfig{figure=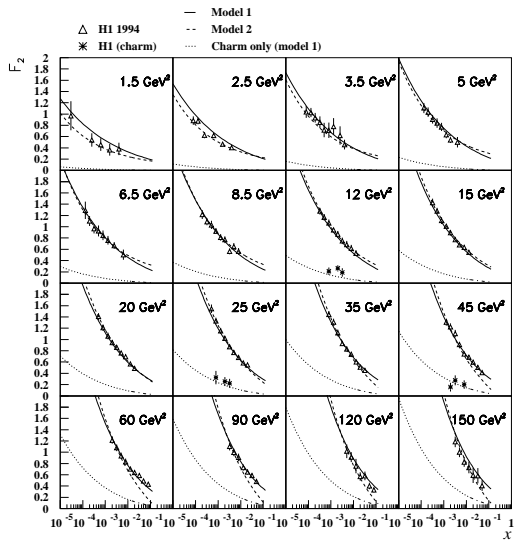,height=7.5cm}}
\vspace{-0.5cm}
\caption{The fits (model 1 and model 2) for the structure function $F_2$.
The structure function $F_2$ and its parametrization are displayed for model 1 and model 2.
dotted line: prediction for the charm component
$F_2^c$, for $M_c=1.5\;\mbox{GeV}$ and model 1. 
The available H1 data are marked by stars.}
\label{fig:fig1}
\end{center}
\end{figure}


\section{Conclusions}

The high energy factorization scheme provides us with some good
predictions for $F_2^c$ which are weakly dependent on the non perturbative
input, within the present error bars on the data.
However, one model predicts rather higher $F_2^c$. More precise data could
help to distinguish between both models.
A good parametrization for the total structure function $F_2$ was obtained 
(3 parameters only are required), but the low value obtained for the
effective coupling constant 
$\alpha_s$ may be an indication of the strong next-leading order BFKL
recently computed \cite{nlbfkl}, which might have been taken into
account effectively in these fits. Anyway, it seems not to spoil the 
$k_\bot$-factorization predictions.

\section*{Acknowledgments}

The material of this contribution is the result of a collaboration with R. Peschanski.
We also thank S. Catani for fruitful suggestions, H. Navelet and Ch. Royon for
stimulating discussions and comments.
This work was supported in part by the EU Fourth Framework Programme `Training 
and Mobility of Researchers', Network `Quantum Chromodynamics and the Deep Structure 
of Elementary Particles', contract FMRX-CT98-0194 (DG 12 - MIHT).

\end{document}